\documentclass[preprint,superscriptaddress,preprintnumbers,aps,amsmath,amssymb]{revtex4}
\usepackage{bm}
\usepackage{color}
\usepackage[pdftex]{graphics}

\usepackage{graphicx}%

\usepackage{multirow}
\usepackage{setspace}
\usepackage{array}
\usepackage{amsmath}
\usepackage{booktabs}
\usepackage{tabularx}
\usepackage{enumerate}
\usepackage{dcolumn}
\usepackage{mathdots}
\usepackage{epstopdf}
 
\usepackage{amsfonts}
\usepackage{float}
\usepackage{subfigure}
\def\prb#1#2#3{{\it Phys.~Rev.}~B~{\bf #1},\ #2\ (#3)}

\def\jcp#1#2#3{J.~Chem.~Phys.~{\bf #1},\ #2\ (#3)}

\def\pra#1#2#3{{\it Phys.~Rev.}~A~{\bf #1},\ #2\ (#3)}

\def\prl#1#2#3{{\it Phys.~Rev.~Lett.}~{\bf #1},\ #2\ (#3)}

\def\rmp#1#2#3{{\it Rev.~Mod.~Phys.}~{\bf #1},\ #2\ (#3)}

\def\k1{k_1}
\def\k2{k_2}
\def\q1{q_1}
\def\q2{q_2}

\def\({\left (}
\def\){\right )}
\def\[{\left [}
\def\]{\right ]}

\newcommand{\beq}{\begin{equation}}
\newcommand{\eeq}{\end{equation}}

\newcount\colveccount
\newcommand*\colvec[1]{
        \global\colveccount#1
        \begin{pmatrix}
        \colvecnext
}
\def\colvecnext#1{
        #1
        \global\advance\colveccount-1
        \ifnum\colveccount>0
                \\
                \expandafter\colvecnext
        \else
                \end{pmatrix}
        \fi
}

\begin{document}
\footnote{?}
\title{
Engineering extended Hubbard models with Zeeman excitations 
of ultracold Dy atoms
}  
\author{R. A. Vargas-Hern\'{a}ndez and R. V. Krems}
\affiliation{Department of Chemistry, University of British Columbia, Vancouver, BC V6T 1Z1, Canada}

\begin{abstract}
We show that Zeeman excitations of ultracold Dy atoms trapped in an optical lattice can be used to engineer extended Hubbard models with tunable inter-site and particle number-non-conserving interactions. 
 We show that the ratio of the hopping amplitude and inter-site interactions in these lattice models can be tuned in a wide range by transferring the atoms to different Zeeman states. 
We propose to use the resulting controllable models for the study of the effects of direct particle interactions and particle number-non-conserving terms on Anderson localization.

\end{abstract}

\maketitle

\clearpage
\newpage

\section{Introduction}

There is currently growing interest in engineering lattice Hamiltonians with ultracold atoms and molecules \cite{quantum-simulations_R}. Of particular interest are extended Hubbard models, which include interactions between particles in different lattice sites. Such models exhibit rich 
physics and have been used to explain the role of long-range interactions in the context of superfluid - Mott insulator transitions \cite{mot}, antiferromagnetism \cite{af, af-2}, high-Tc superconductivity \cite{highTc}, twisted superfluidity \cite{twisted}, supersolids \cite{ss}, self-trapping of bipolarons \cite{john}. Extended Hubbard models are very difficult to solve numerically, especially for two- and three-dimensional lattices. Hence, the need to build experiments, where a many-body quantum system is described by an extended Hubbard model, whose parameters (in particular, the ratio of the hopping amplitude and the inter-site interaction energy) can be tuned by varying external fields, and where the particle densities can be imaged preferably with single site resolution. Tuning the parameters of the model, one could use such experiments to map out the phase diagrams.  
  
There are many proposals for realizing lattice models, including extended Hubbard models 
\cite{PhysModRev-EHM, PhysicaC-EHM, HM-1, zoller-njp, hans-peter, EBHM-magnetic}, with ultracold atoms or molecules trapped in optical lattices. However, if ultracold atoms or molecules are used as probe particles of such models, the inter-site interactions are usually very weak. Therefore,  
the measurements of the phase diagrams require extremely low temperatures and extremely long coherence times, which are often difficult to achieve in current experiments. 
A more promising approach is to trap ultracold molecules in an optical lattice in a Mott insulator phase (with one molecule per site) and use rotational excitations of trapped molecules as probe particles of lattice models 
\cite{NatPhysZoller, Perez-Herrera-Krems, Herrera-Marina-Krems, Gorshkov-Lukin-Rey,Kuns-Rey-Gorshkov,Ping-Marina-Shapiro-Krems, lattice-confined-polar-molecules, Manmana-Rey-Gorshkov, Lukin-Ye-Jin-Rey}. Such excitations can be transferred between molecules in different sites 
due to dipole - dipole interactions. The dynamics of the excitations as well as their interactions can be controlled by external dc electric and/or microwave fields, leading to lattice models with tunable parameters.  
Experiments using excitations as probe particles of lattice models can tolerate much higher temperatures of atomic or molecular motion. 
However, it is currently not possible to create an optical lattice filled uniformly with molecules. On the other hand, ultracold atoms can be trapped in optical lattices with nearly uniform filling \cite{mot, Single-Site-Addressability}. Thus, it would be desirable to engineer extended Hubbard models
with internal excitations of atoms (instead of molecules) trapped in a Mott insulator phase.

A series of experiments have recently demonstrated the cooling of highly magnetic Cr \cite{cr}, Dy \cite{dy-1,dy-2}, and Er \cite{er-1,er-2} atoms to quantum degeneracy. Such atoms interact via long-range magnetic dipole interactions and one can envision engineering the same lattice models with magnetic atoms as with ultracold molecules. However, the internal level structure of magnetic atoms is more complex than the rotational structure of molecules and the nature of magnetic dipole interactions is different from that of electric dipole interactions. 
Motivated by the experiments on magnetic atoms and the work with ultracold molecules, we explore here the possibility of 
engineering extended Hubbard models with internal Zeeman excitations of ultracold magnetic atoms, such as Dy, trapped in a Mott insulator phase.  Exploiting the unique nature of magnetic dipole interactions, we show that, for Zeeman excitations, the ratio of the hopping amplitude and inter-site interaction energy in the resulting lattice models can be tuned in a wide range by transferring the atoms to different Zeeman states. We discuss the advantages of using Zeeman excitations of magnetic atoms over rotational excitations of ultracold molecules. In particular, we show that 
the hopping of the Zeeman excitations in the lattice is insensitive to the magnitude of the magnetic field, which makes the coherent dynamics of excitations robust to field fluctuations. We show that Zeeman excitations in a diluted lattice of Dy atoms undergo Anderson localization over time scales less than one second and propose the models derived here for the study of the role of interactions and particle number fluctuations on Anderson localization.

\newpage
\clearpage

\section{Lattice Hamiltonian with Zeeman excitations}

We consider an ensemble of open-shell atoms with non-zero electron spin ($\bm S$) and orbital angular momentum ($\bm L$) trapped in an optical lattice in the presence of an external DC magnetic field. We assume that the atoms fill the lattice uniformly with one atom per lattice site and that the atoms are not allowed to tunnel between different lattice sites. Thus, the atoms are separated by a large distance ($\ge 260$ nm) equal to half the wavelength of the trapping field. At such separations, the dominant interaction between the atoms in sites $i$ and $j$  is the magnetic dipole - dipole interaction $\hat V_{ij}$.
For simplicity, we assume that the atoms are arranged in a one-dimensional array along the $z$-axis of the space-fixed coordinate frame. In this case,
 \begin{eqnarray}
\hat V_{ij} = \frac{\alpha}{ r_{ij}^3}
\left \{ \frac{1}{2}\left [ \hat J_{i,+} \hat J_{j,-} + \hat J_{i,-} \hat J_{j,+} \right ] - 2 \hat J_{i, z} \hat J_{j,z} \right \}.
\label{eq:Dipol-Dipole_magnetic_2}
\end{eqnarray}
where $\hat J_z$ and $\hat J_{\pm}$ are the $z$-component and the raising/lowering operators of the total angular momentum $\bm J = \bm L + \bm S$, acting on the space of  the 
eigenstates $|J M\rangle$ of $\bm J^2$ and $\hat J_z$, and $\alpha$ is the fine structure constant. 
The full Hamiltonian of the many-atom system is 
\begin{eqnarray}
\hat{\cal{H}} = \sum_i \left \{ A {\bm{L}}_i \cdot {\bm{S}}_i + \mu_{\rm B} ({\bm{L}}_i +  2{\bm{S}}_i)\cdot \bm{ B} \right \} + \frac{1}{2}\sum_i\sum_{j \neq i} \hat{V}_{ij}
\label{eq:Hamiltonian_many-body}
\end{eqnarray}
where  $A$ is the constant of the spin-orbit interaction,  $\mu_{\rm B}$ is the Bohr magneton and $\bm B$ is the vector of an external magnetic field. 

We assume that all atoms are initially prepared in the Zeeman state $| g \rangle$ and a small number of atoms is then transferred to another Zeeman state $| e \rangle$.  Note that the state $|e \rangle$ can be lower or higher in energy than the state $|g \rangle$. 
Following the approach described in Refs. \cite{agranovich} (see also \cite{ping-thesis}), we derive the second-quantized Hamiltonian describing the  Zeeman transitions in this system:
\begin{eqnarray}
{\hat H}_{\rm ex} = v_g + \sum_{i} \sum_{e'} \left \{ \varepsilon_{e'} - \varepsilon_g  + \sum_{j \neq i} \left [  \langle e'_i | \langle g_j | \hat V_{ij} | e'_i \rangle | g_j \rangle -   \langle g_i | \langle g_j | \hat V_{ij} | g_i \rangle | g_j \rangle  \right ] \right \} \hat c^\dagger_{i, e'} \hat c_{i, e'}
\label{outside-1}
\\
+ \sum_{i,j \neq i}  \sum_{e', e''} \langle g_i | \langle e'_j | \hat V_{ij} | e''_i \rangle | g_j \rangle \hat c^\dagger_{i, e''} \hat c_{j, e'} 
+ \sum_{i,j \neq i}  \sum_{e', e''} \left ( 1 - \delta_{e', e''}  \right ) \langle e'_i | \langle g_j | \hat V_{ij} | e''_i \rangle | g_j \rangle \hat c^\dagger_{i, e'} \hat c_{i, e''} 
\label{outside-2}
\\
 + \frac{1}{2} \sum_{i,j \neq i}  \sum_{e', e''} \sum_{f', f''} \left [ \delta_{e', e''} \delta_{f', f''} \langle g_i | \langle g_j | \hat V_{ij} | g_i \rangle | g_j \rangle + \langle e'_i | \langle f'_j | \hat V_{ij} | e''_i \rangle | f''_j \rangle \right.
\nonumber
 \\
\left. - 2 \delta_{f', f''} \langle e'_i | \langle g_j | \hat V_{ij} | e''_i \rangle | g_j \rangle \right ] \hat c^\dagger_{i, e'} \hat c_{i, e''} \hat c^\dagger_{j, f'} \hat c_{j, f''}
\label{outside-3}
\\
+\sum_{i, j \neq i} \sum_{e'} \left [ \langle g_i | \langle g_j | \hat V_{ij} | g_i \rangle | e'_j \rangle \hat c_{j, e'}    +  \langle g_i | \langle e'_j | \hat V_{ij} | g_i \rangle | g_j \rangle \hat c^\dagger_{j, e'}    \right ]
\label{outside-4}
\\
+ \frac{1}{2} \sum_{i,j \neq i}  \sum_{e', e''} \left [  \langle g_i | \langle g_j | \hat V_{ij} | e'_i \rangle | e''_j \rangle \hat c_{i, e'} \hat c_{j, e''}  + \langle e'_i | \langle e''_j | \hat V_{ij} | g_i \rangle | g_j \rangle \hat c^\dagger_{i, e'} \hat c^\dagger_{j, e''}    \right ] 
\label{outside-5}
\\
+ \frac{1}{2} \sum_{i,j \neq i}  \sum_{e', e'', f'} \left [  \langle e'_i | \langle g_j | \hat V_{ij} | e''_i \rangle | f'_j \rangle   - \delta_{e',e''}   \langle g_i | \langle g_j | \hat V_{ij} | g_i \rangle | f'_j \rangle \right ]
 \hat c^\dagger_{i, e'} \hat c_{i, e''} \hat c_{j, f'} 
\label{outside-6}
\\
+ \frac{1}{2} \sum_{i,j \neq i}  \sum_{e', e'', f'} \left [  \langle e'_i | \langle f'_j | \hat V_{ij} | e''_i \rangle | g_j \rangle   - \delta_{e',e''}   \langle g_i | \langle f'_j | \hat V_{ij} | g_i \rangle | g_j \rangle \right ]
 \hat c^\dagger_{i, e'} \hat c_{i, e''} \hat c^\dagger_{j, f'} 
\label{outside-7}
\end{eqnarray}
where
\begin{eqnarray}
v_g =  N \varepsilon_g + \frac{1}{2} \sum_{i} \sum_{j \neq i} V_{ij}^{gg},   
\end{eqnarray}
$N$ is the number of atoms, $\varepsilon_g$ and $\varepsilon_{e'}$ are the energies of the atomic states $|g \rangle$ and $|e'\rangle$, and
\begin{eqnarray}
V_{ij}^{gg} = \langle g_i | \langle g_j | \hat V_{ij} | g_i \rangle | g_j\rangle.
\label{vgg}
\end{eqnarray}  
Here, we assume that the Zeeman states $e', e'', f', f'' \neq g$ and use the operators $\hat c^\dagger_{i,e'}$  and $\hat c_{i,e'}$ defined by $\hat c^\dagger_{i, e'} |g_j \rangle = \delta_{ij} | e'_j \rangle$ and $\hat c_{i, e'} |e'_j \rangle = \delta_{ij} | g_j \rangle$.
For the purposes of this work, it is convenient to rewrite this complex Hamiltonian as  
\begin{eqnarray}
\hat{H}_{\rm ex} =  v_g + \sum_{i} (\Delta \varepsilon _{eg} + d_i)\hat{c}^{\dagger}_{i}\hat{c}_{i} + \sum_i \sum_{j\neq i} t_{ij}  \hat{c}^{\dagger}_{j}\hat{c}_{i} + \hspace{1.cm}
\label{tight-binding}
\\
\frac{1}{2}\sum_i \sum_{j \neq i} v_{ij} c^\dagger_i c_i c^\dagger_j c_j + \hspace{3.cm}
\label{exciton-interactions}
\\
 \frac{1}{2} \sum_i \sum_{j\neq i} t_{ij}  \left ( \hat{c}^{\dagger}_{i}\hat{c}^\dagger_{j} + \hat{c}_{i}\hat{c}_{j} \right )+ 
\sum_i \sum_{j\neq i} s_{ij}  \left ( \hat{c}^{\dagger}_{i} + \hat{c}_{i} \right ) + 
\sum_i \sum_{j\neq i} p_{ij} \left (  \hat{c}^{\dagger}_{i} +   \hat{c}_{i} \right ) \hat{c}^\dagger_j \hat{c}_j
\label{nonconserving}
\\
+ {\cal H}(e' \neq e, g) \hspace{4.cm}
\label{outside-space}
\end{eqnarray}
where the operators $\hat c^\dagger_i$  and $\hat c_i$ are defined by $\hat c^\dagger_i |g_j \rangle = \delta_{ij} | e_j \rangle$ and $\hat c_i |e_j \rangle = \delta_{ij} | g_j \rangle$,
 $\Delta \varepsilon _{eg}$ is the energy separation between the states $|e \rangle$ and $| g \rangle$, and the parameters of the Hamiltonian are
\begin{eqnarray}
d_i = \sum_{j\neq i} d_{ij},
\label{di}
\end{eqnarray}
\begin{eqnarray}
d_{ij} = \left \{V_{ij}^{ge} - V_{ij}^{gg} \right \},
\label{dij}
\end{eqnarray}
\begin{eqnarray}
v_{ij} = V^{ee}_{ij} + V^{gg}_{ij} - 2 V^{eg}_{ij},
\label{interaction-constant}
\end{eqnarray}
\begin{eqnarray}
V_{ij}^{eg} = V_{ij}^{ge} = \langle g_i | \langle e_j | \hat V_{ij} | g_i \rangle | e_j\rangle
\label{vge}
\\
V_{ij}^{ee} = \langle e_i | \langle e_j | \hat V_{ij} | e_i \rangle | e_j\rangle
\label{vee}
\end{eqnarray}
\begin{eqnarray}
t_{ij} =  \langle g_i | \langle e_j  | \hat V_{ij}  | e_i \rangle  | g_j \rangle
\label{j-amplitude}
\\
s_{ij} = \langle e_i | \langle g_j | \hat V_{ij} | g_i \rangle | g_j\rangle 
\label{s-amplitude}
\end{eqnarray}
and
\begin{eqnarray}
p_{ij} = \langle e_i | \langle g_j | \hat V_{ij} | e_i \rangle | e_j\rangle  -
\langle e_i | \langle g_j | \hat V_{ij} | g_i \rangle | g_j\rangle.
\label{p-amplitude}
\end{eqnarray}

The terms (\ref{tight-binding}), (\ref{exciton-interactions}) and (\ref{nonconserving}) are a part of the full Hamiltonian that describes the Zeeman transitions only within the four-state subspace $|a\rangle |b\rangle$ with both $|a\rangle$  and $|b\rangle$ being either $|g \rangle$ or $|e\rangle$.   If the energy gap for the $|g \rangle \rightarrow |e\rangle$ transition were far detuned from all other energy gaps in the Zeeman level spectrum, it would be sufficient to consider the part of the Hamiltonian given by Eqs. (\ref{tight-binding}), (\ref{exciton-interactions}) and (\ref{nonconserving}). 
It is important to note that for highly magnetic atoms it may be necessary to consider Zeeman states outside of this subspace. Figure 1a shows that the Zeeman states of a Dy atom in the ground electronic state form a ladder of nearly equidistant levels at weak magnetic fields.  This pattern of energy levels is characteristic of highly magnetic atoms with zero or negligible hyperfine structure. This pattern of energy levels allows for transitions to states outside of the subspace spanned by  $|g \rangle$ or $|e\rangle$. For example, two atoms in the $|g\rangle$ state may interact to produce two Zeeman states with energies just above and just below that of $|g\rangle$.  Such interactions are induced by the matrix elements in Eq. (\ref{outside-5}). 
The full Hamiltonian must also include the terms that describe the interactions of two atoms in states $e', e'' \neq g$ to produce atoms in other states $f', f'' \neq g, e$. Since the majority of atoms are in a particular state $|g \rangle$, we assume that such interactions are unlikely and neglect them. 

Various lattice models can be engineered by controlling the magnitude of the different matrix elements of the magnetic dipole interaction entering Eqs. (\ref{outside-1}) - (\ref{outside-7}). 

\section{Engineering lattice models}

In this section we show (i) how to simplify the lattice Hamiltonian presented in Section II by applying magnetic fields; and (ii) how to tune the relative magnitudes of the parameters of the resulting lattice models by transferring atoms into different states. We illustrate the tunable range of the parameters by calculating the model parameters for the specific example of Dy atoms in an optical lattice.

\subsection{$t-V$ model}

Eqs. (\ref{tight-binding}) and (\ref{exciton-interactions}) represents a $t-V$ model \cite{t-V},  an extended single band, Hubbard model for hard-core bosons \cite{zoller-njp, extended_hubbard_R,phase-diagram}.
This model can be studied with the Zeeman excitations if the effect of the terms (\ref{nonconserving}) and (\ref{outside-space}) are suppressed. As we show below, this can be achieved by applying a finite magnetic field and introducing a small admixture of different $M$-states into the eigenstates $|J M\rangle$. 

Eqs. (\ref{outside-1}) - (\ref{outside-7}) and (\ref{tight-binding}) -- (\ref{nonconserving}) can be separated into terms that conserve the number of excitations (Eqs. \ref{outside-1} -- \ref{outside-3}, \ref{tight-binding} and \ref{exciton-interactions}) as well as particle number-non-conserving terms (Eqs. \ref{outside-4} -- \ref{outside-7} and \ref{nonconserving}). If the Zeeman states form a ladder of equidistant states, the particle number-non-conserving terms can be further separated into  energy-conserving (some terms in Eq. \ref{outside-5}) and energy-non-conserving terms (Eqs. \ref{outside-4} -- \ref{outside-7}, \ref{nonconserving}). The effect of the energy-non-conserving terms can be eliminated by applying a finite magnetic field such that the energy difference between the Zeeman levels is significantly larger than the magnitude of the matrix elements appearing in Eqs. (\ref{outside-4}) -- (\ref{outside-7}) and (\ref{nonconserving}). 

In order to eliminate the effect of all terms in Eq. (\ref{outside-space}), it is necessary to make the energy gap for the $|g\rangle \rightarrow |e\rangle$ transition unique, i.e. different from the energy gaps in the Zeeman spectrum just below and just above the states $|g\rangle$ and $|e\rangle$. This can be achieved by applying a magnetic field strong enough to shift the Zeeman levels due to couplings between different total angular momentum states. 
As illustrated in the lower panel of Figure 1, these couplings introduce a differential in the energy gaps between different Zeeman states.
To illustrate this, we plot in Figure 1b the of the energy gaps between the states correlating with the states $|J=8, M=-1\rangle$ and $|J=8, M=0\rangle$; states $|J=8, M=0\rangle$ and $|J=8, M=+1\rangle$ and states $|J=8, M=+1\rangle$ and $|J=8, M=+2\rangle$, as functions of $B_0$.
As Figure 1b shows, the magnetic field with $B_0 \approx 200-300$ G produces the differential of the energy gaps equal to the matrix elements $t_{i, i+1}$ for Dy atoms on an optical lattice with $a = 266$ nm. At fields with $B_0 > 300$ G, the difference in the energy gaps becomes much larger than any of the matrix elements in Eq. (\ref{outside-space}) so the Hamiltonian (\ref{tight-binding}) -- (\ref{outside-space}) reduces to the $t-V$ model.

The parameters of the $t-V$ model can be tuned by transferring atoms into different Zeeman states. 
If the $|g \rangle$ and $|e\rangle$ states are the Zeeman states $|g\rangle = |J M\rangle$ and $|e \rangle = |JM'\rangle$, the matrix elements (\ref{vgg}) and (\ref{j-amplitude}) of the operator (\ref{eq:Dipol-Dipole_magnetic_2}) can be written as follows:

\begin{eqnarray}
d_{ij} = V_{ij}^{ge} - V_{ij}^{gg} =   \frac{2\alpha}{r_{ij}^3} \left (M^2 - M'M \right )
\label{magnetic-Dij}
\end{eqnarray}
and 
\begin{eqnarray}
t_{ij} = \frac{\alpha}{2r^3_{ij}}   \left [ a^i_+ b^j_-\delta^i_{M',M+1} \delta^j_{M',M-1} + a^i_- b^j_+\delta^i_{M',M-1} \delta^j_{M',M+1} \right ],
\label{eq:Dipol-Dipole_magnetic_matrix elements}
\label{magnetic-Jij}
\end{eqnarray}
with 
\begin{eqnarray}
a^i_\pm = \left [ J(J+1) - M(M \pm 1) \right ]^{1/2} \\
b^j_\pm = \left [ J(J+1) - M'(M' \pm 1) \right ]^{1/2}
\end{eqnarray}
The interaction between the Zeeman excitations (\ref{interaction-constant}) can be written as
\begin{eqnarray}
v_{ij} = - \left [ (V_{ij}^{eg} - V_{ij}^{gg}) +  ( V_{ij}^{eg} - V_{ij}^{ee}) \right ] =  - \frac{2\alpha}{r_{ij}^3} \left ( M  - {M'}\right)^2
\label{magnetic-Vij}
\end{eqnarray}

These equations show that the diagonal matrix elements $V_{ij}^{gg}$ and $V_{ij}^{eg}$, and hence $d_{ij}$ and $v_{ij}$ are non-zero, provided both $M \neq 0$ and $M' \neq 0$. 
This is different from the case of the electric dipole - dipole interaction between molecules \cite{herrera-pra}. The electric dipole interaction must couple states of the opposite parity. Therefore, if $|g\rangle$ and $|e\rangle$ are the eigenstates of a molecular Hamiltonian in the absence of electric fields, the matrix elements $d_{ij}$ and $v_{ij}$ of the electric dipole - dipole interaction vanish. These interactions can be induced in an ensemble of polar molecules by applying an external electric field that mixes the rotational states with different parity \cite{herrera-pra, ping}


In contrast, the matrix elements of the magnetic dipole - dipole interaction (\ref{magnetic-Dij}) and (\ref{magnetic-Jij}) should not be expected to vary significantly with an external magnetic field. This will be illustrated and discussed in the following section, using the example of Dy atoms on an optical lattice. 
As follows from Eqs.  (\ref{magnetic-Dij}) and (\ref{magnetic-Jij}), the relative weights of the two couplings can be tuned by choosing different Zeeman states $|J M \rangle$ as the $|g \rangle$ and $|e \rangle$ states. Notice, for example, that for the particular case of $|g \rangle$ being the state $|J, M=0\rangle$, the magnitudes of $d_{ij}$, and consequently $d_i$, vanish.




\subsection{$t-V$ model with Dy atoms}

We illustrate the range of controllability of the parameters of the $t-V$ models using an example of Dy atoms in an optical lattice. 
The absolute magnitudes of $d_{ij}$, $t_{ij}$ and $v_{ij}$ increase with $J$ as the square of the magnetic moment.  
The ground electronic state of Dy is characterized by the total angular momentum $J=8$ so Dy atoms have a large magnetic moment (10 Bohr magnetons) and a manifold of Zeeman states displayed in Figure 1a. The Zeeman structure of Dy allows for the possibility of using the state $|M=0\rangle$ as the $|g\rangle$ state, leading to the value $d_{ij} = 0$.

If the states for the Zeeman excitations in an ensemble of Dy atoms are chosen to be well-defined angular momentum states  $|g\rangle = |J M\rangle$ and $|e\rangle = |J M' \rangle$, Eq. (\ref{magnetic-Jij}) shows that $t_{ij} = 0$ unless $|M-M'| = 1$. Eq. (\ref{magnetic-Vij}) shows that the interaction $v_{ij}$ 
is $ \propto (M-M')^2$ so it is independent of $M$ and $M'$, if $|M-M'| = 1$. However, the parameter $t_{ij}$ is sensitive to the magnitudes of $M$ and $M'$. This is illustrated in the upper panel of Figure 2. 
The ratio $t_{ij}/v_{ij}$ can thus be tuned by transferring atoms into the Zeeman states with different $M$, as illustrated in the lower panel of Figure 2. Notice that the ratio $t_{ij}/v_{ij}$ is always negative, which means that the interactions between the excitations are always effectively attractive. The largest magnitude of the ratio $t_{ij}/v_{ij} \approx -18$ can be achieved when the atoms are prepared in the Zeeman state with $M=0$ and excited to the Zeeman state with $M=+1$, while the smallest magnitude of the ratio $t_{ij}/v_{ij} \approx -4$ can be achieved by preparing the atoms in the maximally stretched state $|J=8, M=-8\rangle$ or $|J=8, M=+8\rangle$. 

As illustrated in Figure 4 the absolute magnitude of $v_{ij}$ can be tuned if the atoms are prepared in coherent superpositions of 
states with different $M$. Consider for example the superpositions $|g \rangle = \alpha | J M\rangle + \beta |J, M + \delta \rangle$ and $|e \rangle = \alpha' | J M'\rangle + \beta' |J, M' + \delta' \rangle$. 
For the parameter $t_{ij}$ to be non-zero, either $|M-M'|$ or $|M-M' + \delta - \delta'|$ must be 1. 
However, there is no such restriction on the matrix elements determining the magnitude of $v_{ij}$. As follows from Eq. (\ref{magnetic-Vij}), the magnitude of $v_{ij}$ is expected to increase with increasing the difference between the angular momentum projections of the states participating in the excitation. 
This is graphically illustrated in Figure 3, showing that the magnitude of $v_{ij}$ can reach 600 Hz, if $M-M' = 16$. This suggests that the ratio $t_{ij}/v_{ij}$ can be tuned by preparing the atoms in the coherent superpositions of the following kind: $|g \rangle = \alpha | J M\rangle + \beta |J, M + \delta \rangle$ and $|e \rangle = \alpha' | J M + 1\rangle + \beta' |J, M + \delta' \rangle$. The parameters $t_{ij}$
and $v_{ij}$ for these states are both non-zero and the magnitude of $v_{ij}$ can be modified by varying the value of $|\delta - \delta'|$.

\begin{figure}[ht]
\label{figure1}
\begin{center}
\includegraphics[scale=1.00]{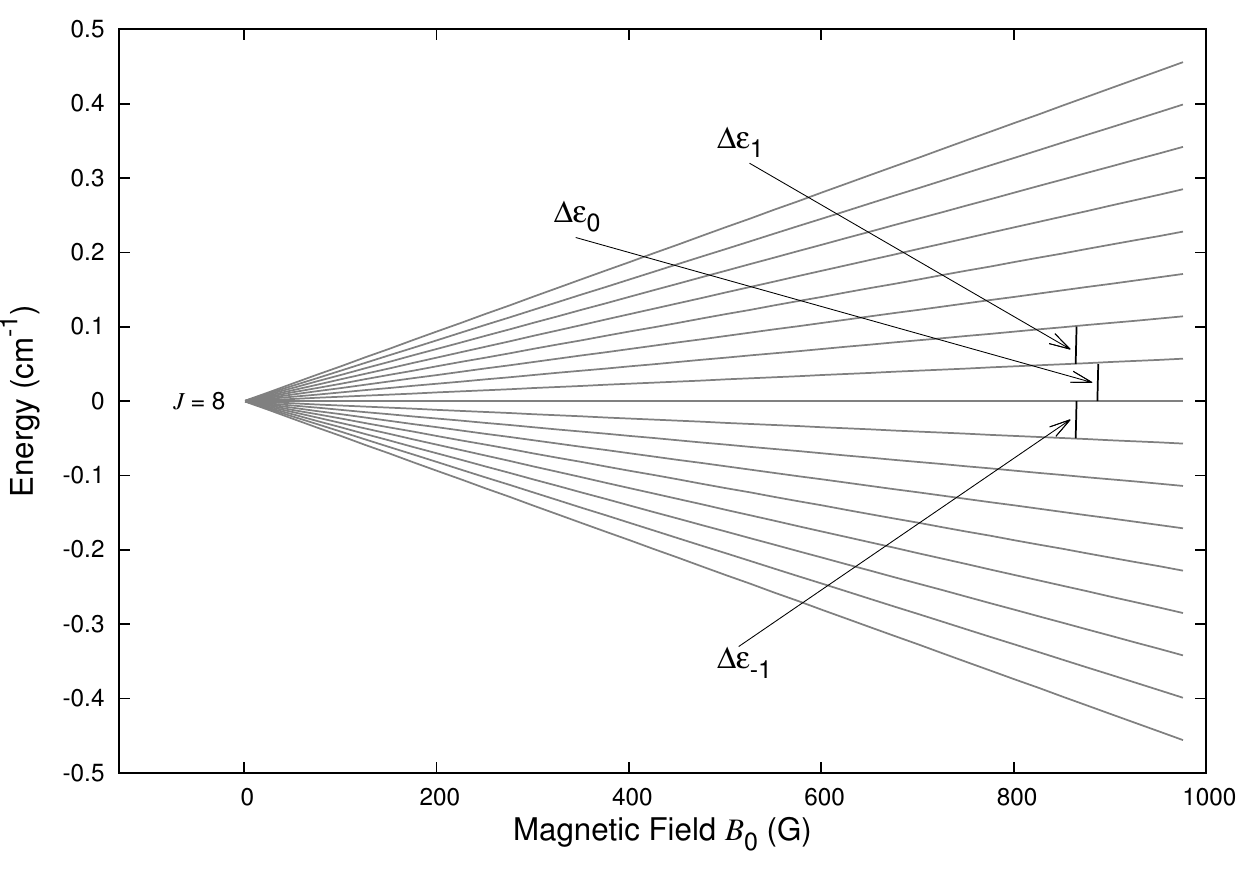} \\
\includegraphics[scale=1.00]{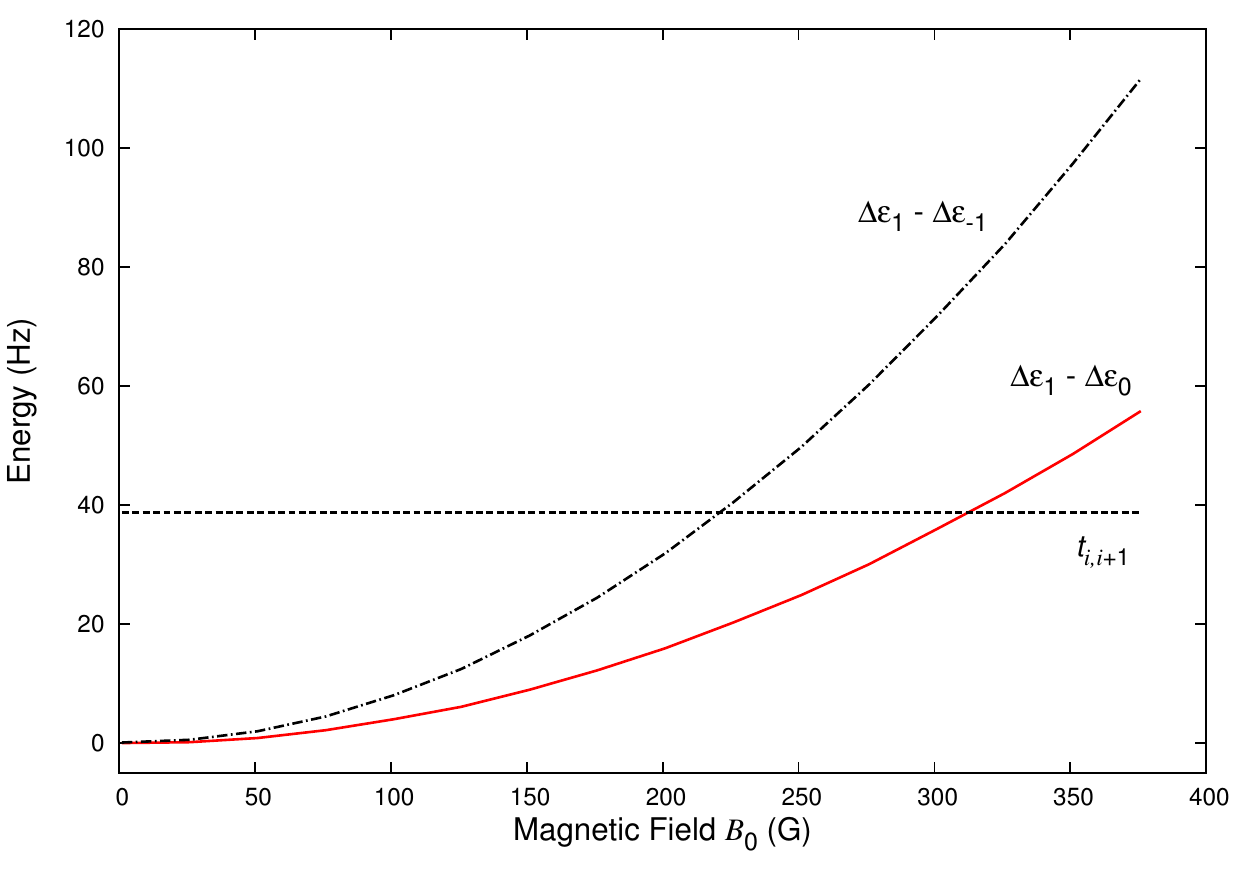} 
\end{center}
\caption{Upper panel: Zeeman levels of a Dy($^5I$) atom in the lowest-energy spin-orbit state characterized by $J = 8$ in a magnetic field $\bm B = B_0 \hat {z}$. Lower panel: the solid curve -- difference of the energy gaps $(\varepsilon_{M=2} - \varepsilon_{M=1}) - (\varepsilon_{M=1} - \varepsilon_{M=0})$; 
the dot-dashed curve -- difference of the energy gaps $(\varepsilon_{M=2} - \varepsilon_{M=1}) - (\varepsilon_{M=0} - \varepsilon_{M=-1})$. 
The horizontal dashed line shows the magnitude of the matrix element $t_{i,i+1}$ in Eq. (\ref{j-amplitude}) for Dy atoms with $|g \rangle = |J=8, M=0 \rangle$ and $|e \rangle = |J=8, M=1 \rangle$ in an optical lattice with $a = 266$ nm.
}
\end{figure}

\begin{figure}[ht]
\label{figure2}
\begin{center}
\includegraphics[scale=0.7]{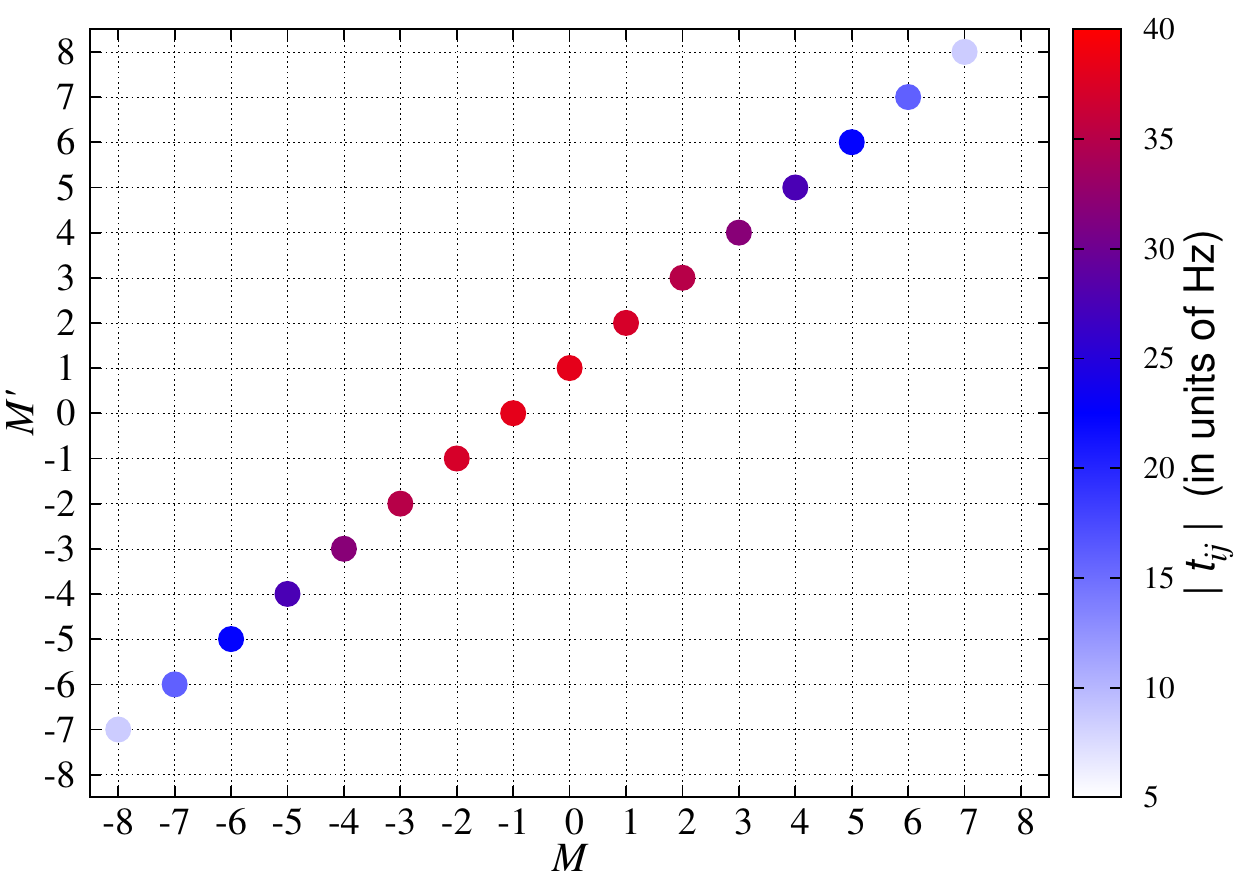} \\
\includegraphics[scale=0.7]{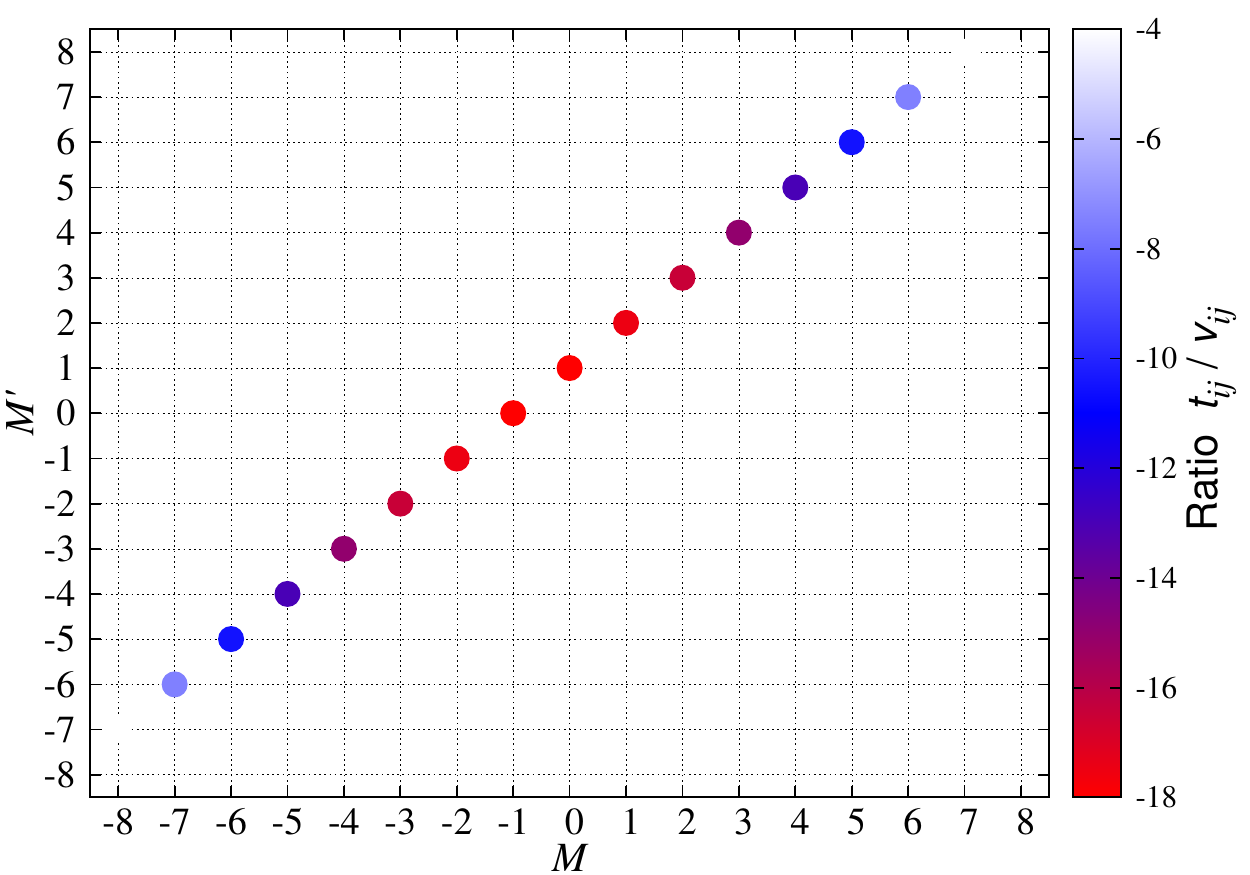} 
\end{center}
\caption{The magnitudes of the coupling constants $t_{ij}$ (upper panel) and the ratio $t_{ij}/v_{ij}$ (lower panel) with $j = i \pm 1$ for the Zeeman states of Dy corresponding to $|g \rangle \Rightarrow | J M\rangle$ and $|e \rangle \Rightarrow | J M'\rangle$. The calculations are for the magnetic field $\bm B = B_0 \left (0.1 \hat x + \hat z \right )$ with $B_0 = 100$ G. 
The Zeeman states in this magnetic field retain $96 \%$ of the eigenstates of $\bm J^2$ and $\bm J_z$. 
}
\end{figure}

\begin{figure}[ht]
\label{figure3}
\begin{center}
\includegraphics[scale=1.0]{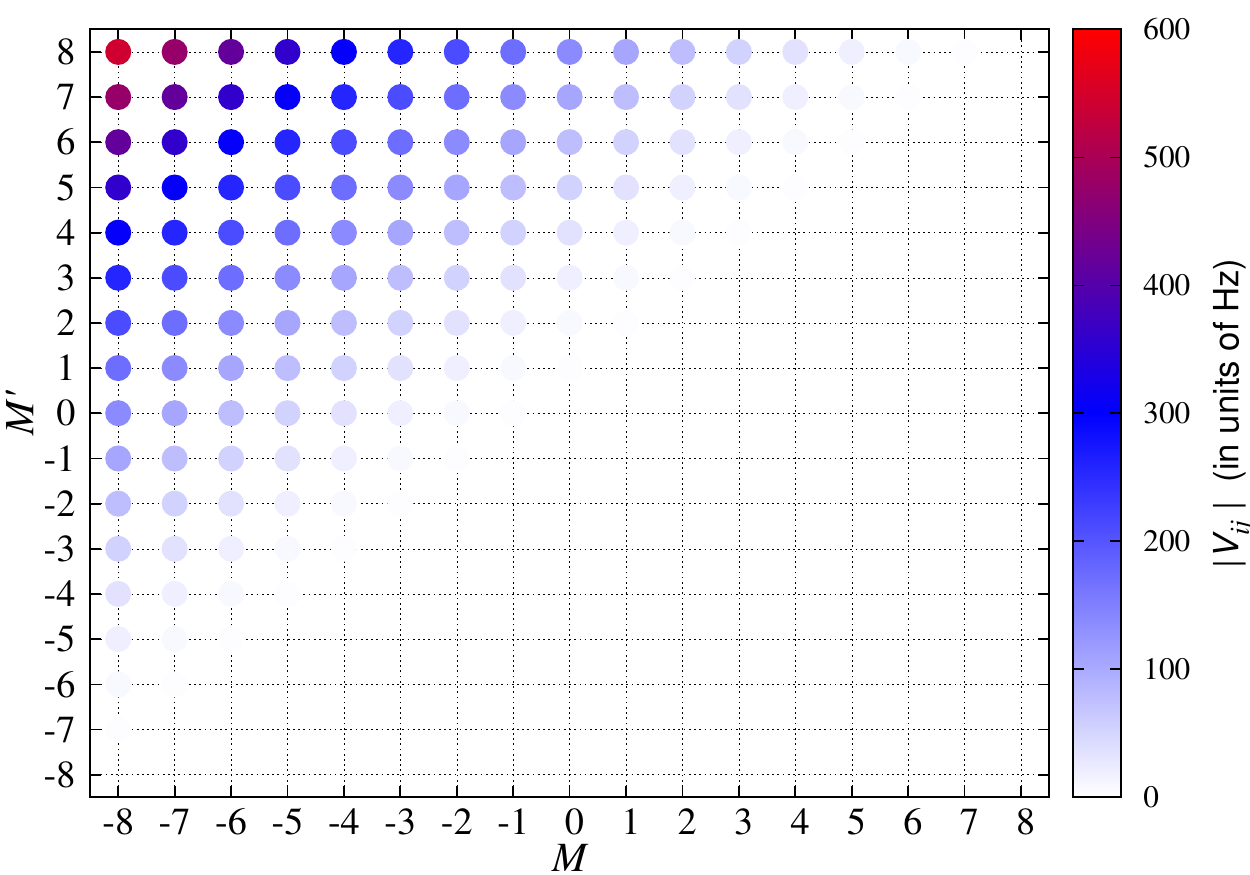} 
\end{center}
\caption{The magnitude of the coupling constant $v_{ij}$ with $j = i \pm 1$ for the Zeeman states of Dy corresponding to $|g \rangle \Rightarrow | J M\rangle$ and $|e \rangle \Rightarrow | J M'\rangle$. 
The calculations are for the magnetic field $\bm B = B_0 \left (0.1 \hat x + \hat z \right )$ with $B_0 = 100$ G. 
The Zeeman states in this magnetic field retain $96 \%$ of the eigenstates of $\bm J^2$ and $\bm J_z$.  }
\end{figure}

\begin{figure}[ht]
\label{figure4}
\begin{center}
\includegraphics[scale=1.00]{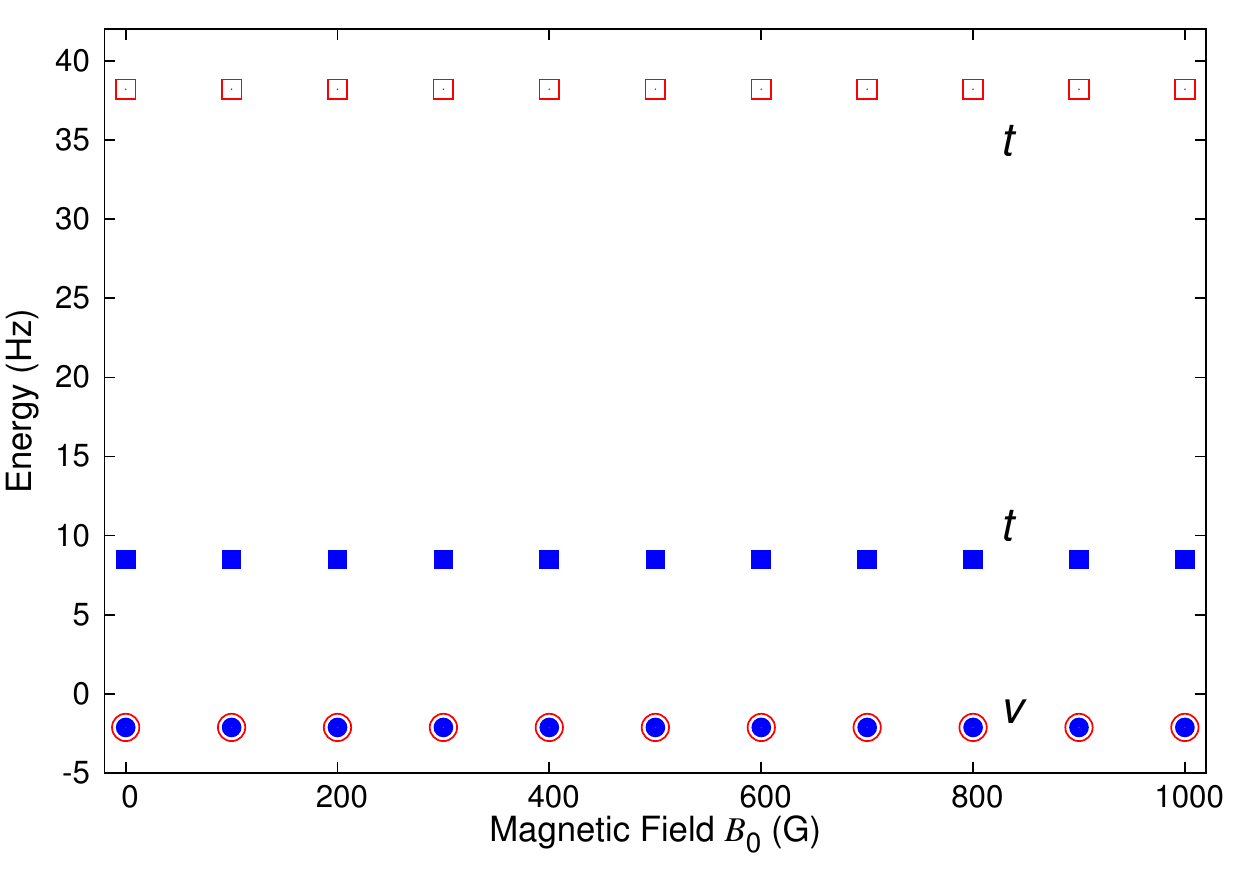}
\end{center}
\caption{The magnetic field dependence of  the quantities $t_{i,i+1}$ (squares) and $v_{i,i+1}$ (circles) defined in Eqs. (\ref{magnetic-Dij}) and (\ref{magnetic-Jij}) for two different pairs of the Zeeman state
of Dy($J=8$) atoms:
the full symbols -- the results for $ | g  \rangle =  | J=8,M=-8  \rangle$ and $ | e  \rangle =  | J=8,M=-7  \rangle$;
the open symbols -- the results for $ | g  \rangle =  | J=8,M=0  \rangle$ and $ | e  \rangle =  | J=8,M=+1  \rangle$.
The magnetic field is given by $\bm B = B_0 \left (0.1 \hat x + \hat z \right )$. 
The Zeeman states in such a magnetic field retain $96 \%$ of the eigenstates of $\bm J^2$ and $\bm J_z$. 
}
\end{figure}

The interaction of atoms with a magnetic field couples states with different total angular momenta $J$, which may - in principle - modify the atomic states $|g\rangle$ and $|e\rangle$, and, consequently, the lattice model parameters.
It is important to examine the effect of an external magnetic field on the lattice model parameters. 
To do this we diagonalized the full Hamiltonian of the Dy atom in a magnetic field $\bm B = B_0 \left (0.1 \hat x + \hat z \right )$ and used the eigenstates to evaluate the model parameters in Eqs. (\ref{tight-binding}) -- (\ref{exciton-interactions}). Since the states of different $J$ in the Dy atom are separated by large energy gaps ($> 1000$ cm$^{-1}$) due to the spin-orbit interaction,  the eigenstates of Dy in a magnetic field are nearly identical to the angular momentum states $|J M\rangle$. Figure 5 shows the nearest-neighbour coupling parameters $t_{i,i+1}$ and $v_{i,i+1}$ for a one-dimensional array of Dy atoms on an optical lattice with the lattice site separation $a = 266$ nm computed for two pairs of Zeeman states at different magnetic fields. The results shown in Figure 4 illustrate that the Hamiltonian parameters do not change with the magnetic field in the interval of field strengths between zero and 5000 G. 
 This is important because it shows that the magnetic field can be used to separate the Zeeman states in order to create isolated two-level systems or tuned to the limit of vanishing field where the terms in Eq. (\ref{nonconserving}) become important, without affecting the parameters of excitation interactions.

\begin{figure}[ht]
\label{figure5}
\begin{center}
\includegraphics[scale=0.9]{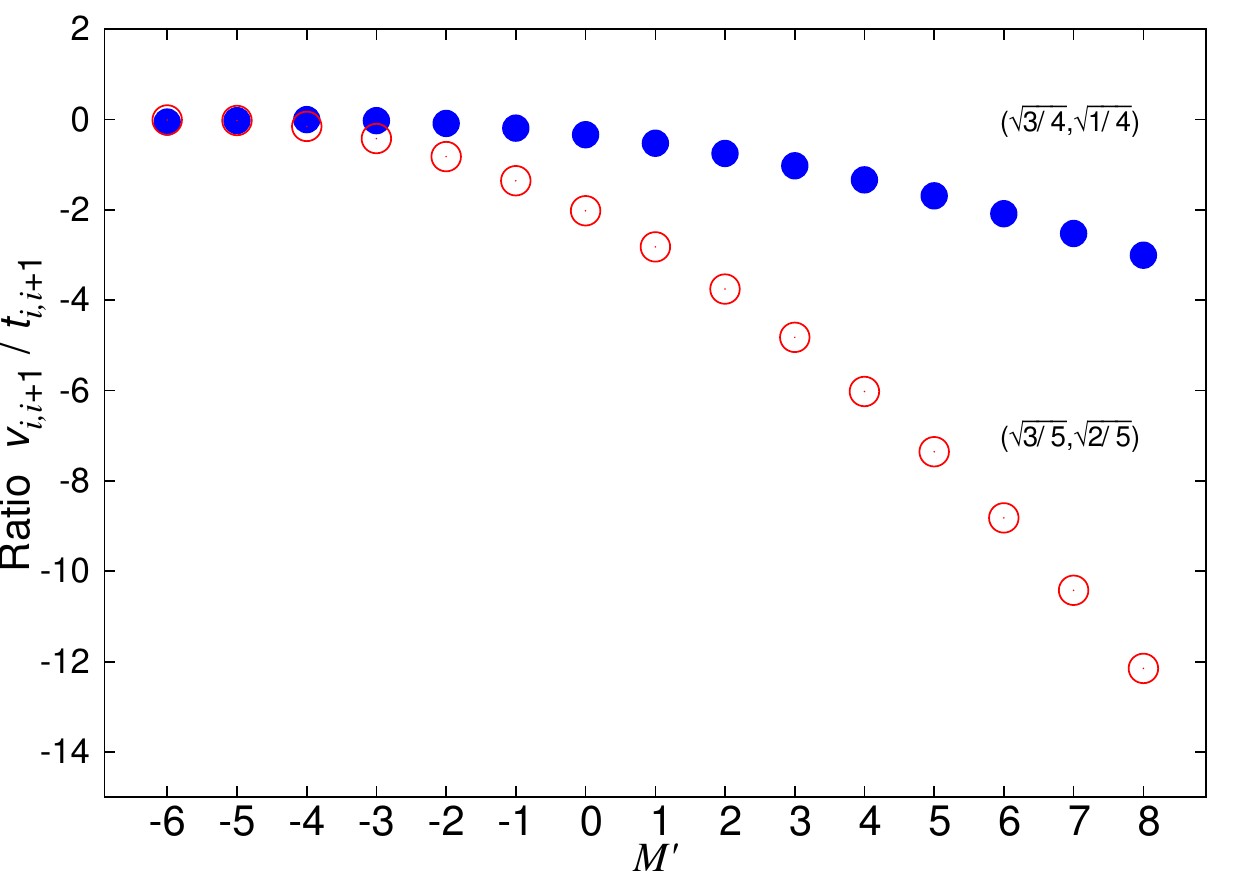}
\end{center}
\caption{The ratio $v_{i,i+1}/t_{i,i+1}$ for the Zeeman states of Dy corresponding to $ | g  \rangle =  | J=8,M=-7  \rangle$ and 
$ | e  \rangle = a  | J=8,M=-8  \rangle + b |J=8, M'\rangle$: full circles -- $a=\sqrt{3/4}, b = \sqrt{1/4}$; open circles -- $a= \sqrt{3/5}, b = \sqrt{2/5}$.
The calculations are for the magnetic field $\bm B = B_0 \left (0.1 \hat x + \hat z \right )$ with $B_0 = 100$ G. 
The Zeeman states in this magnetic field retain $96 \%$ of the eigenstates of $\bm J^2$ and $\bm J_z$. 
}
\end{figure}

\begin{figure}[ht]
\label{figure7}
\begin{center}
\includegraphics[scale=0.7]{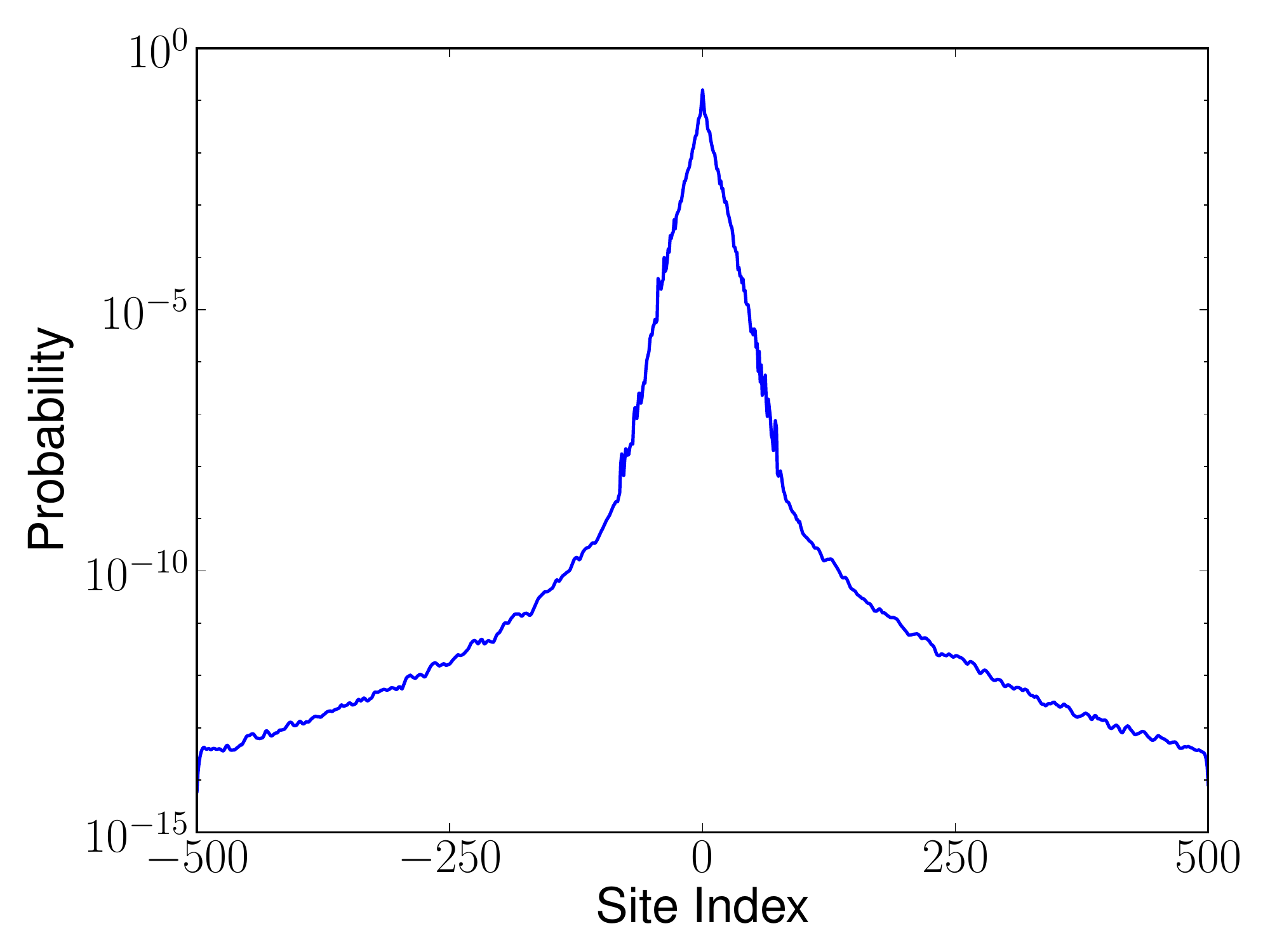}
\includegraphics[scale=0.7]{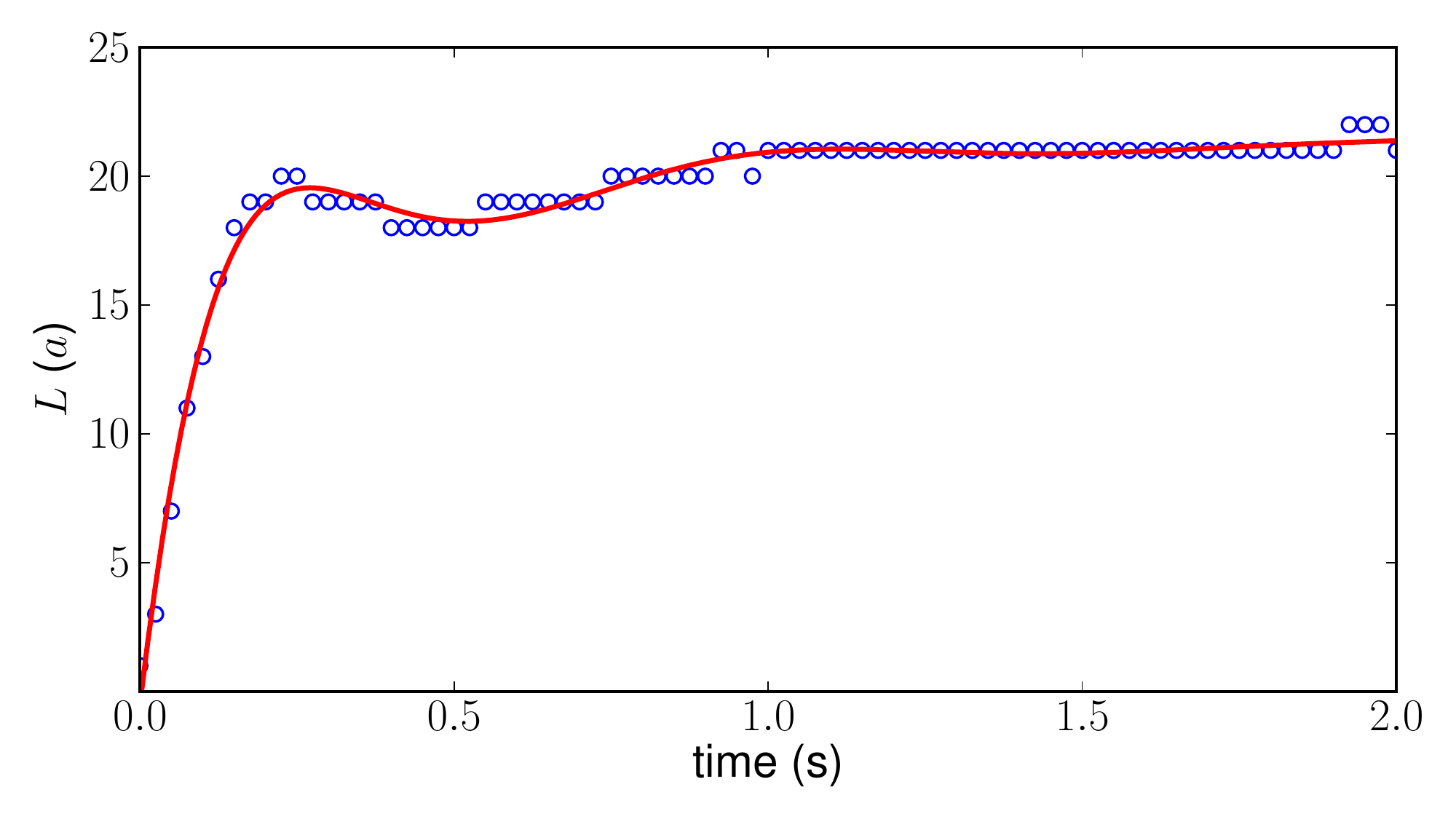}
\end{center}
\caption{Anderson localization of the $|J=8, M=0 \rangle \rightarrow |J=8, M=+1 \rangle$ excitation in a one-dimensional array of Dy atoms on an optical lattice with $a=266$ nm and 20 \% of the lattice sites empty. The upper panel shows the probability distribution for the atoms in the corresponding site to be in the excited state at $t=2$ seconds formed by a single excitation placed at $t=0$ in the middle of a lattice with 1000 sites.  The lower panel shows the width of the excitation probability distirbution as a function of time. 
}
\end{figure}

\subsection{Particle number-non-conserving interactions}

In the limit of weak magnetic fields, as $\Delta \varepsilon_{eg} \rightarrow 0$, the energy separation between different particle number states of the model (\ref{tight-binding})
decreases to the minimum of $d_i$. As follows from Eq. (\ref{magnetic-Dij}), this parameter can be eliminated if the ground state $|g\rangle$ is chosen to be $|J,M=0\rangle$. 
At weak magnetic fields, the particle number-non-conserving terms (\ref{nonconserving}) as well as the terms in Eq. (\ref{outside-space}) must be included in the Hamiltonian. Number non-conserving interactions may mediate effective long-range hopping (for example, a particle can move in a lattice by virtual transitions to the three-particle subspace and back). As such, these interactions may have non-trivial effects on the dynamics of quantum walks and localization in disordered lattices. Such interactions arise in the context of excitons in molecular crystals \cite{agranovich-nnc}. However, they are usually negligibly small and difficult to investigate. As shown below, number non-conserving interactions can be made significant in the system considered here. 


We first note that if the array of atoms is arranged along the magnetic field direction, the matrix elements of the operator (\ref{eq:Dipol-Dipole_magnetic_2}) that determine the parameters $s_{ij}$ and $p_{ij}$ in Eq. (\ref{nonconserving}) vanish. This simplifies the resulting lattice models to include only the first of the particle number-non-conserving terms in Eq. (\ref{nonconserving}). 
If desired, the terms $s_{ij}$ and $p_{ij}$ can be tuned to finite values if the magnetic field direction is changed or the atoms are prepared in coherent superpositions of different $M$-states. For example, if $|g\rangle = |J,M=0\rangle$ and $|e\rangle = \alpha |J=8,M=0\rangle + \beta |J=8,M=1\rangle $, all of $t_{ij}$, $s_{ij}$ and $p_{ij}$ become non-zero. 
Here, we assume that the magnetic field is directed along the atomic array and that
$s_{ij}=0$ and $p_{ij}=0$.

Care must be taken when considering the limit $\Delta \varepsilon_{eg} \rightarrow 0$.
In this limit, multiple Zeeman states become degenerate and it may be necessary to consider interband couplings determined by Eq. (\ref{outside-space}). This may be useful if complicated models, including multiple excitations of different kind, are desired.  Note, however, that  if $|g\rangle$ and $|e\rangle$ are states with well-defined $M$ and $M'$, a two-atom state $|M\rangle |M' \rangle$ can only be coupled to the same state, the state $|M'\rangle |M \rangle$ or a state $|M \pm 1\rangle |M' \mp 1 \rangle$.   The matrix elements of the dipole - dipole interaction $\langle M, M' | \hat V_{ij} | M \pm 1, M' \mp 1 \rangle$ change the number and type of excitations in the atomic ensemble. These processes can be eliminated if the state $|g \rangle$ is chosen to be $| J, M = \pm J \rangle$. In this case, the effective lattice model describing the dynamics of $|g \rangle \rightarrow |e\rangle$ excitations is

\begin{eqnarray}
\hat{H}_{\rm ex} =  v_g + \sum_{i} (\Delta \varepsilon _{eg} + d_i)\hat{c}^{\dagger}_{i}\hat{c}_{i} + \sum_i \sum_{j\neq i} t_{ij}  \hat{c}^{\dagger}_{j}\hat{c}_{i} + \hspace{1.cm}
\nonumber
\\
\frac{1}{2}\sum_i \sum_{j \neq i} v_{ij} c^\dagger_i c_i c^\dagger_j c_j + 
 \frac{1}{2} \sum_i \sum_{j\neq i} t_{ij}  \left ( \hat{c}^{\dagger}_{i}\hat{c}^\dagger_{j} + \hat{c}_{i}\hat{c}_{j} \right )
\label{kitaev-interactions}
\end{eqnarray}

It is important to note that this model is valid as long as $\Delta \varepsilon _{eg}$ (which is determined by the magnitude of the magnetic field) is significantly larger than $t_{ij}$. In this limit, the effect of the number-non-conserving terms is perturbative, i.e. a single excitation remains predominantly in the single-particle subspace, undergoing virtual transitions to the three-particle subspace. If the energy gap $\Delta \varepsilon _{eg}$ is so small that the interactions (\ref{kitaev-interactions}) lead to the creation of multiple particles, other terms in Eq. (\ref{outside-5}) must be included, making the Hamiltonian more complex.

If the effects of the interactions $v_{ij}$ are to be removed, one can choose the states $|g\rangle = |J, M=0\rangle$ and $|e\rangle = |J, M=1\rangle$. In this case, $|t_{ij}| \gg |v_{ij}|$ (see Figure 4). However, the lattice model for these excitations is also affected by terms in Eq. (\ref{outside-5}), which lead to leaking of the $|e \rangle$-state populations to other Zeeman states of higher energy. These terms lead to the spontaneous creation of atoms in Zeeman states above and below the energy of the state with $M = 0$, as well as the inverse process. 
The Zeeman state populations must eventually return to states $e$ and $g$, as the total number of the Zeeman states is finite and small. These terms thus serve as an additional source of particle number-non-conserving interactions that generate atoms in state $e$.

\subsection{Anderson localization of Zeeman excitations}

Until now, we assumed that the atoms populate the optical lattice uniformly. If the lattice is populated partially (which is more often the case in experiments), the empty lattice sites serve as impurities that can scatter the Zeeman excitations. Since the distribution of empty sites is random, the Zeeman excitations thus propagate in a randomly diluted lattice. Tuning the models as described above suggests an interesting opportunity to explore the role of direct particle interactions and number non-conserving forces on Anderson localization in disordered lattices \cite{anderson,review-rmp}. In addition, the ability to design optical lattices with various dimensionalities and geometries can be used to verify the scaling hypothesis of Anderson localization \cite{scaling} as well as Anderson localization of particles with long-range hopping in various geometries \cite{long-range-AL}. 
Here, we explore if the parameters of the models based on Zeeman excitations of Dy are significant enough to allow Anderson localization over experimentally feasible time- and length-scales.

We consider an isolated Zeeman excitation in a one-dimensional array of 1000 Dy atoms trapped in an optical lattice with $a=266$ nm containing 20 \% of empty lattice sites. We use the parameters correpsonding to the $|J=8, M=0 \rangle \rightarrow |J=8, M=+1 \rangle$ excitation and compute the dynamics of quantum walk for the Zeeman excitation placed at $t=0$ on a single atom in the middle of the lattice. The wave packet of the excitations is propagated by computing the time-evolution operator, as described in detail in Ref. \cite{tianrui}. The results of each dynamical propagation are averaged over 100 disorder realizations (random distributions of empty lattice sites).

The results shown in Figure 6 illustrate that the Zeeman excitation forms an exponentially localized spatial distribution within one second. The width of the distribution characterized as the length $L$ containing 90 \% of the excitation probability exhibits a short-time oscillation which is likely an effect of coherent back scattering and approaches the value of $\sim 20$ lattice sites in the limit of long time. These results can be directly mapped onto the results describing Anderson localization for rotational excitations in an ensemble of polar molecules \cite{tianrui} and the electronic excitations in an ensemble of Rydberg atoms \cite{francis}. 

\section{Conclusion}

In this work, we consider Zeeman excitations in an ensemble of  highly magnetic atoms (such as Dy) trapped in an optical lattice, with one atom per lattice site. The Zeeman excitations can travel in the lattice due to energy transfer between the atoms. The most important results of this work can be summarized as follows:  

\begin{itemize}
\item
We show that superpositions of the Zeeman excitations can be used to simulate the $t-V$ model (the single-band, extended Bose-Hubbard model for hard-core bosons). The parameters of the model (most importantly, the ratio of the hopping amplitude and the inter-site interaction energy) can be tuned by preparing  the atoms in different Zeeman states. For an ensemble of Dy atoms on an optical lattice with $a= 266$ nm, we show 
that the inter-site interaction can be engineered to be as large as 600 Hz. 

\item
We illustrate that the parameters of the model (hopping amplitudes and inter-site interactions) are insensitive to the magnetic field. This has two significant consequences. 
First, an external magnetic field can be used to uncouple the electron degrees of freedom from nuclear spins, thereby removing complications associated with the hyperfine structure of atoms and the degeneracies of the Zeeman states. Second, an external magnetic field can be used to separate the Zeeman states, leading to suppression of energy- and particle number-non-conserving terms.

\item
We show that the same Hamiltonian can be used to simulate a lattice model with significant $c^\dagger_i c^\dagger_j$ terms, leading to particle number interactions. These interactions mediate effective interactions modifying the hopping of particles and can be used to produce entangled pairs \cite{felipe-nnc}. 

\item Since the lattice with randomly distributed empty sites leads to a quantum percolation model for the Zeeman excitations, we propose to apply the models derived here for the study of Anderson localization induced by off-diagonal disorder. In particular, our results suggest the possibility of studying the role of inter-site interactions and particle number fluctuations on quantum localization in diluted lattices. We show that for an optical lattice with $a= 266$ nm partially populated with Dy atoms, Anderson localization of excitations placed on individual atoms occurs over timescales less than a second. 

\end{itemize}

\section*{Acknowledgment}
We thank Tianrui Xu for the calculations presented in Figure 6 and John Sous for useful discussions. The work is supported by NSERC of Canada.

\end{document}